\documentclass{article}
\usepackage{a4wide}
\usepackage{lineno,hyperref}
\usepackage{listings}
\usepackage{graphicx}
\usepackage[printonlyused]{acronym}
\begin{document}
	
\title{On-line Remote {\sc ekg} as a Web Service}

\author{Augusto Ciuffoletti\thanks{Dept. of Computer Science, Universit\'a di Pisa L.go B. Pontecorvo - 56122 Pisa ORCID:0000-0002-9734-2044 } \\ e.mail: augusto.ciuffoletti@unipi.it}

\maketitle

\begin{abstract}
	A 3-leads, non-diagnostic {\sc ekg} has a role in emergency rescue and homecare. In this paper we introduce the design and a prototype of a service, provided to a doctor and a patient, for the on-line remote visualization of patient's 3-leads {\sc ekg}. The architecture is based on the {\sc http} protocol, using commercial off-the-shelf devices to implement the sensor on patient's side, a browser on a laptop \ac{PC} on the doctor's side as viewer, and a cloud container to connect the two using Websockets. A prototype is built to evaluate signal latency, power consumption of the patient side device, and the quality of the rendering. After some experiments, latency is measured below $1sec$, and power consumption is estimated in the 2A*3.3V range; visualization is comparable to commercial, non-diagnostic products. The prototype patient device is portable, and can be operated using rechargeable battery packs. Its cost is below 100\$, and all the required equipment is commercially available. The architecture is ready for {\em on field} evaluation, and we indicate how to improve power consumption while reducing cost.
\end{abstract}

\section{Introduction}

There are cases where heartbeat monitoring would improve doctor's assistance, but the patient and the physician are not in the same room. What we need in that case is to transmit the \ac{EKG} from patient's heart to doctor's display: this is an {\em on-line remote \ac{EKG}}. The number of use cases for remote \ac{EKG} is long and includes assistance in rural areas, emergency rescue, automated processing and alert, long term recording etc. In most cases a diagnostic 12-leads measurement is inappropriate, since it requires time and a specific training to be prepared, while a 3-leads, non-diagnostic \ac{EKG} is fitting. 
Two stories illustrate its possible utilization.

One is that of an ambulance with paramedical personnel rescuing a injured person after a car accident. It is likely that the location is covered by a ground or satellite broadband provider, and that an \ac{AP} is available, e.g. by tethering a smartphone. With an on-line remote \ac{EKG} service the 3-leads trace is delivered to the medical staff in the hospital, that defines the emergency level.

A different homecare story is that of a cardiac patient that is periodically contacted by the family doctor to check general conditions. In that case the physicist uses a remote \ac{EKG} service to receive the trace without leaving its desk during a phone call, possibly interacting with the patient about electrodes position or body posture.

This paper addresses the long distance delivery of a non-diagnostic \ac{EKG}, using cloud facilities integrated with personal devices.

\section{Background and previous works}

The electrical functionality of the heart is a fundamental diagnostic information for the medical staff. Its recording dates back to the end of the 19th century. From that time, research addressed first the amplification of the signal, whose amplitude is in the $\mu V$ range, and next the design of filters that remove unwanted components (like powerline {\em hum}, or signals from other muscular activity). The current state of the art on \ac{EKG} filtering is in \cite{har16a}.

Computers come into play in connection with filters, and today they extract the fundamental features of the \ac{EKG} signal. In \cite{gut15a} the authors introduce an approach and evaluate the computing resources needed, with a survey of other research results in the field. A recent survey is also in \cite{nai16a}. 

With the advent of the Web, storage and transport of \ac{EKG} data emerge as a practical perspective.
In \cite{lug05a} the author investigates its use for human identification, and incidentally creates a database of \ac{EKG}s, that later became a precious resource for many. The availability of historical data is relevant for medical research, daily health-care use, and educational purposes.

The primary concern with the transport and storage of an \ac{EKG} is about its confidential nature.
In \cite{sat17a} the authors survey a number of cryptographic systems that can be used to secure the transport and the storage of biosensor data. The paper assumes the presence of a number of distinct sensing devices aggregated in a \ac{MSN}.

The ubiquitous presence of wireless networks justifies the realization of portable (or even {\em wearable}\cite{bai17a}) \ac{EKG} devices that forward the trace to a nearby \ac{PC} or tablet. When many such devices are aggregated in a network, we speak of a \ac{BAN}. In 2001, \cite{jon01a} introduces the basic principles and concepts, and, ten years later, a survey in \cite{che11a} lists nine \ac{BAN} projects, 5 of which include \ac{EKG}. Detailed designs of \ac{BAN} devices are found, for instance, in (\cite{shn05a}), \cite{ple10a}, \cite{cri15a}. However they are appropriate for local area networks, confined in a room or a building. 

The availability of cloud servers allows overcoming such limits. In \cite{xia13a} the authors propose a cloud-based infrastructure: the physician and the patient are clients of a Web server that processes the \ac{EKG}, evaluates its quality, and provides parameter extraction and visualization. The server is in the \ac{AWS} cloud, while the client uses personal devices that interact with the server using \ac{HTTPS}.

To facilitate the diffusion of the remote \ac{EKG} service, we need to address the cost of the hardware devices, which also meets the needs of developing countries and of rural areas, as in \cite{wal09a} and \cite{deb17a}.

In this paper we show how a cloud infrastructure can be used to relay the \ac{EKG} from the patient's device to the doctor's laptop: in a nutshell, we actualize the ideas in \cite{xia13a}, but with the target of \cite{jon01a}. This result is obtained using low cost devices and established protocols and infrastructures, without overlooking security aspects.

\section{An open service for remote {\sc ekg}\label{sec:method}}

\begin{figure*}
\centering
\includegraphics[width=0.9\linewidth]{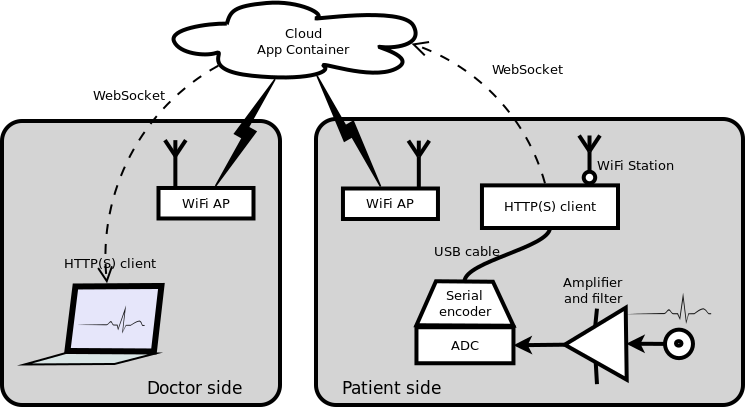}
\caption{Remote {\sc ekg} service infrastructure layout}
\label{fig:outline}
\end{figure*}

The design guidelines authorize several alternatives, and we propose the architecture outlined in figure \ref{fig:outline}.

The right box represents the devices on patient's side. We observe the analog stage with the operational amplifier and the filters, the processor that encodes the analog signal on a serial line, a \ac{HTTP} user agent that further processes the signal and manages the Websocket. This stage is connected to the Internet with a wireless link, and reaches the cloud server.

The devices on doctor's premises are in the left box: a wireless \ac{AP}, and the smart device (a \ac{PC} or smartphone) with a \ac{HTTP} client that manages the Websocket, and displays the \ac{EKG} in a convenient way.

From the security perspective, we highlight that only the routers inside the \ac{AP}s need to expose a public \ac{IP} address, therefore both patient's and doctor's devices can be protected by a \ac{NAT} and other security techniques implemented on the router, according with common practice.

The cloud server needs a limited computational capacity, and it may exist just for the time needed by the physicist to examine the \ac{EKG}: it is the ideal candidate for a microservice instance \cite{new15a}. A \ac{PaaS} provider can create a new instance of a microservice in a few seconds, make it accessible in the Internet with a unique \ac{URL}, and destroy it after use.

The cloud provider is responsible for the security of the server, whose \ac{URL} is known only to the partners. However, server life span is so limited that the potential intruder has little time to identify and attack it, especially if communication is encrypted, for instance with \ac{TLS}.

The adoption of \ac{HTTP} as the transport protocol is a cornerstone, since cloud services, that are reachable only using this protocol, provide a flexible, reliable and secure infrastructure for \ac{EKG} transmission. Although real time aspects of the remote \ac{EKG} are incompatible with the {\sc http} request/response mechanism, the {\em WebSockets} \cite{rfc6455} have been recently introduced to allow unsolicited communication between client and server. WebSockets have a behavior similar to \ac{TCP} channels, but are encapsulated in a \ac{HTTP} session.

In our architecture, the \ac{PaaS} server acts as a relay point: two separate WebSockets are created to this end, one with the patient, another with the doctor, and the data are trasparently transferred from the former to the latter. This configuration turns out to be easily scalable, with costs that at the prototype stage can be null, and with excellent security features.

In addition, with \ac{HTTP} the doctor uses the browser of his laptop to view the \ac{EKG}, without the need to install new software. The architecture is therefore agnostic of the operating system installed on doctor's premises.

To have similar benefits also on patient's side, we introduce the following principles, that, in this paper, are synthesized in the \ac{OpLoC} acronym:
\begin{itemize}
	\label{txt:principles}
	\item {\bf Open Source}: components and protocols are exhaustively documented and freely reproducible;
	\item {\bf Low Cost}: the less expensive option is always preferable. In particular, if a functionality is already available, it is not re-implemented;
	\item {\bf Commercial Devices}: devices must be available on retail (aka \ac{COTS}). 
\end{itemize}

Put together, the three principles ensure that an implementation is easily reproducible, and makes a solid ground for further investigation. In addition, low cost ensures that its applicability is not selective by scale and wealth.

\section{A prototype for a remote \ac{EKG} service}

\begin{figure*}[h]
\centering
\includegraphics[width=0.7\linewidth]{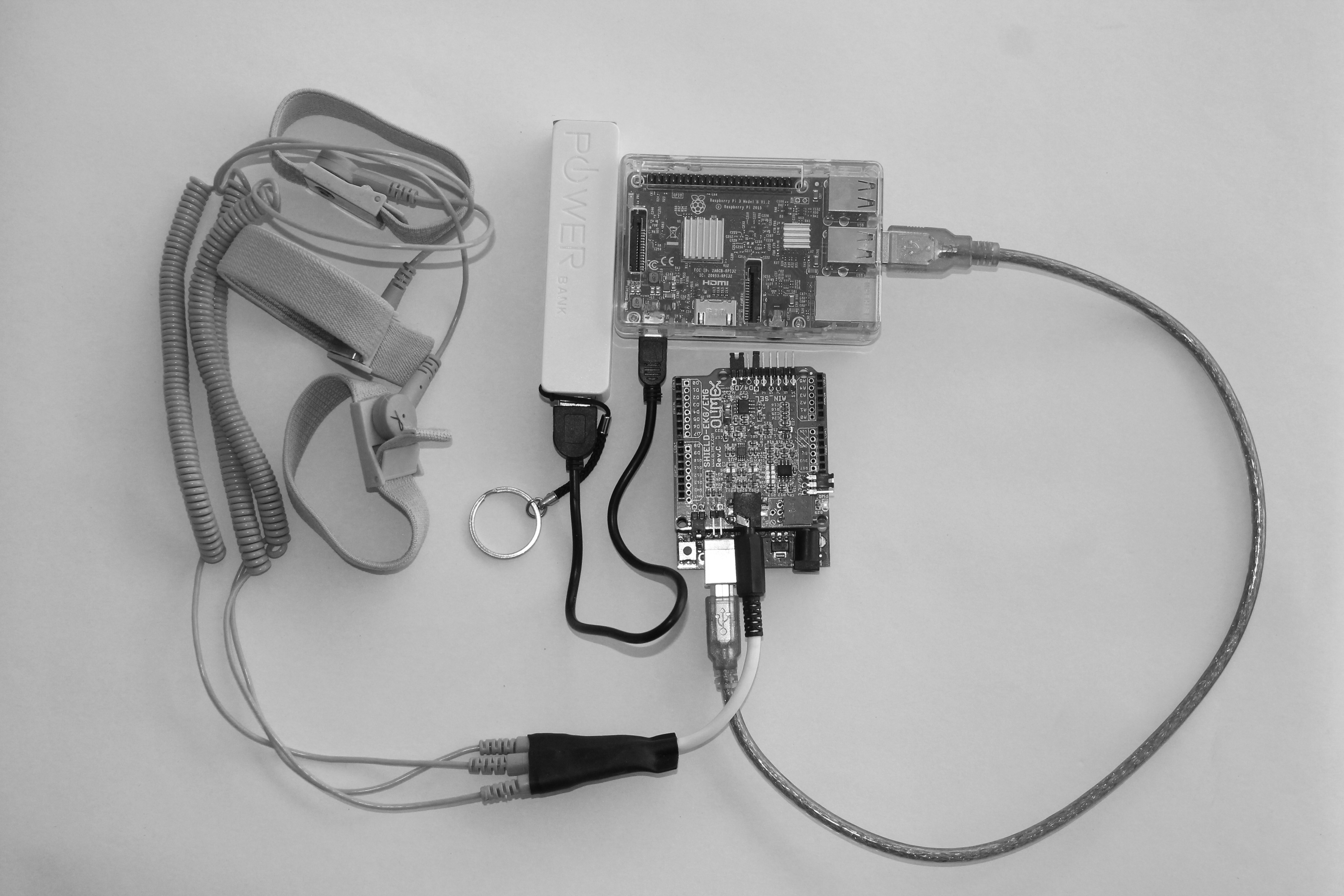}
\caption{The patient-side prototype device, with three elastic bands, the $2200 mAh$ battery, the Arduino/Olimex {\sc ekg} boards and the Raspberry Pi 3}
\label{fig:patient}
\end{figure*}

This section describes a concrete implementation of the above abstract design. In figure \ref{fig:patient} we see the device on patient's side, while in figure \ref{fig:doctor} we see the display in the browser.
Together with the \ac{OpLoC} hardware components, the prototype contains also three ad-hoc software products, whose code is publicly available on the Bitbucket and GitHub platforms:
\begin{itemize}
	\item the analog-to-serial encoder in the \ac{MCU} \cite{cur:olimex}
	\item the {\sc http} patient-side \ac{UA} \cite{cur:ecgClient}
	\item the doctor's page and the Websocket manager in the cloud container \cite{cur:ecgServer}
\end{itemize}

\subsection*{The sensor: amplifier and filter}

The prototype uses the popular {\em Olimex EMG-EKG} sensor board \cite{olimexEkgEmg}: its size is $6\times8\times3cm$, and the cost is around 30\$. It is usually sold with \ac{EKG} pads.

The board integrates a 3rd order filter at 40Hz and two high-pass filters to remove high frequencies and baseline drift.

A single board acquires a non-diagnostic 3-lead \ac{EKG} through a 3-pole jack. The design of the board allows to stack up to 6 boards for a diagnostic 12-lead \ac{EKG}, but our prototype uses just one of them.

\subsection*{The \ac{MCU}: \ac{ADC} and serial encoder} 

The \ac{MCU} is a popular Arduino Uno board, whose cost is around 10\$. The connection with the Arduino-compliant Olimex board described above is obtained by {\em stacking} the two boards: the result is mechanically stable and sufficiently compact. The analog output of the sensor board is converted into a 10-bits integer by the \ac{ADC} embedded in the \ac{MCU}.

\begin{table}
%\lstinputlisting[language=c++, 
%	firstline=52, 
%	lastline=65,
%	basicstyle=\linespread{1}\footnotesize]{
%		./ecg-sketch/ecg-sketch.ino}
\begin{verbatim}
void Timer2OveflowISR( ) {
  int i;
  unsigned long int t;
  if ( full [b] ) {
    Serial.println (”fail”);
    return;
  }
  for (int Channel = 0 ; Channel < 6 ; Channel++) {
    Data [ Channel ] [ b ] = analogRead (Channel );
  }
  full[b]= true;
  b=bˆ1;
}
\end{verbatim}
\caption{The function triggered every $4msecs$ on the {\sc mcu} \label{code:interrupt}}
\end{table}

To have an accurate timing, needed for filtering and analysis purposes, a hardware interrupt triggers data fetch with a frequency of 250Hz \cite{piz85a}. The code snippet for the interrupt handler is in table \ref{code:interrupt}: note that it is prepared to collect six analog data for a standard 12-lead \ac{EKG}, but only one is used in the prototype. The main loop waits for the buffer to be readable, and encodes a space-separated line:

\begin{verbatim}
<h>:<m>:<s>.<msec> <v1> <v2> <v3> <v4> <v5> <v6>\n
\end{verbatim}

The first field is a $1 msec$ resolution timestamp, and the other six fields, of which only one is used in the prototype, are integers in the interval $[0-1023]$ that correspond to a sample.
The encoding does not introduce any rounding or information loss, and its redundancy is functional to data integrity. The timestamp is not as accurate as the sampling period: it is used only for rendering.

To avoid interference between data fetch and encoding, a two positions buffer is introduced, with access regulated by a semaphore. 

It is worth saying that Olimex provides a different driver (\cite{olimexEkgEmg}), but its accuracy is not sufficient for the task.

\subsection*{\ac{MCU} to \ac{SBC} interface}

The two units communicate using their \ac{UART}s, with a baudrate of 115200. Since the maximum length of a line is 44 8-bit characters, and 250 lines are produced each second, the \ac{MCU} outputs 11000 bytes per second. Since one stop bit per byte is added, this requires a baudrate of 99000, which is consistent with \ac{UART}s one.
  
\subsection*{Patient side user-agent}

Since the sensor unit has tight timing requirements and limited capabilities, it is more appropriate to decouple the \ac{HTTP} \ac{UA} on a distinct platform. The {Raspberry Pi 3} used in the prototype is a \ac{SBC} that uses production-level libraries to implement the patient-side user agent. Its cost is about 25\$. 

The \ac{UA} is written in Python, and opens a Websocket connection with a container hosted in the Heroku cloud whose \ac{URL} is:
\begin{verbatim}
http://<container fqdn>/in/<id>
\end{verbatim}

where \verb|<id>| is a unique id for the device. When the Websocket is opened, the \ac{UA} starts sending \ac{EKG} samples encoded as \ac{JSON} objects with two fields: the timestamp, and one value.

\subsection{The server}

The server is a container in the Heroku cloud: for the sake of simplicity, the prototype envisions one single container hosting several concurrent \ac{EKG}s instead of a short lived micro-service for each of them.

The server is implemented using the {\em Python/Flask} micro-framework and the {\tt gunicorn} \ac{WSGI} \ac{HTTP} server. It provides three families of application routes:
\begin{itemize}
	\item {\tt /} which returns a presentation page,
	\item {\tt /in/<id>} used by a patient's \ac{UA} to open a Websocket, as explained above,
	\item {\tt /<id>} used by doctor's \ac{UA} to open a Websocket and receive the \ac{EKG},
	\item {\tt /out/<id>} used as endpoint for doctor's Websocket
\end{itemize} 

The server waits for a patient's \ac{UA} request on the \verb|/in/<id>| route, upgrades the session to a Websocket and prepares to receive a request from doctor's side with the same \verb|<id>|: until then, all received data are lost. If doctor's request arrives first, the server returns a negative reply.

When the server receives doctor's request on an \verb|/<id>| path matching patient's \ac{UA}, it delivers a page containing the JavaScript code to open the WebSocket on \verb|/out/<id>| and displays the \ac{EKG}.

Since each patient \ac{UA} has a different identifier, a single server may host several \ac{EKG} at the same time, each of which is received by only one doctor \ac{UA}.

The security mechanisms announced in the \ref{sec:method} are not fully implemented: namely, in the prototype the server is persistent and communications use plain \ac{HTTP}. The implementation of such features appears to be quite straightforward, and it is a matter for the improvement of the product.  

\subsection*{The display \ac{UA}}

The \ac{UA} on doctor's side is a web browser running on a laptop. The doctor opens the \ac{HTTP} session using a \ac{URL} with the \ac{FQDN} of the server and the \verb|/<id>| route. If the corresponding patient is already connected, the response contains a resource composed of the familiar \ac{EKG} canvas and of a JavaScript application that opens a WebSocket on the \verb|/in/<id>| route. From it the \ac{EKG} is received and the Chart.js library is used to display it.

The JavaScript application filters out the $50 Hz$ powerline noise with a moving average of five values and finds R peaks to compute heartbeat frequency. This demonstrates the possibility of additional filtering and feature extraction on doctor's side, and provides a readable result.

\begin{figure*}[h]
\centering
\includegraphics[width=0.9\linewidth]{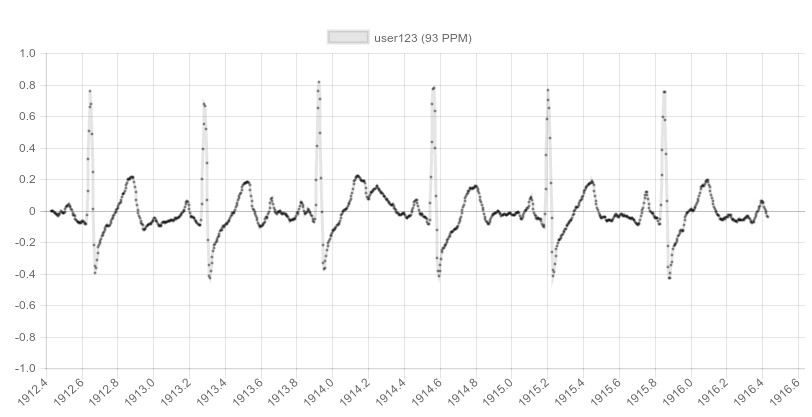}
\includegraphics[width=0.9\linewidth]{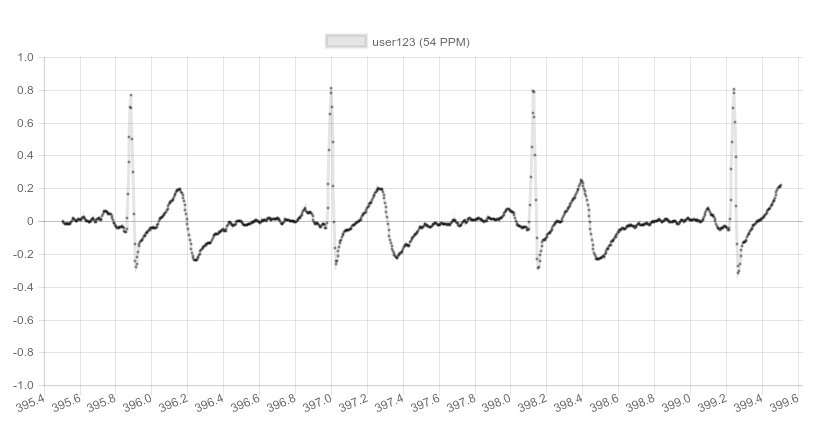}
\caption{
\label{fig:ekg_run_b}Two \ac{EKG}s: one after exercise, using elastic bands on left and right wrists and right ankle, another at rest, using adhesive electrodes on chest. They have been captured as screenshots form doctor's browser using the prototype described in the paper.}
\end{figure*}

\section{Results and future work}

A relevant parameter is the latency between the production of the signal and its visualization: it has been measured as lower than one second. Autonomy with battery operation has an impact on practical usability as well: using a $2200 mAh$ rechargeable power bank an autonomy of 72 minutes has been measured, and the prototype operates with the 3.3V power supply provided by the Raspberry.

The Raspberry Pi 3 over-kills its task, and it is a candidate for replacement with a more focused \ac{OpLoC} device. Being in charge of implementing the patient's WebSocket, it uses 50\% of the space, 90\% of the power (nearly 1500 mA), and 30\% of the cost.

\begin{figure*}[h]
	\centering
	\includegraphics[width=0.7\linewidth]{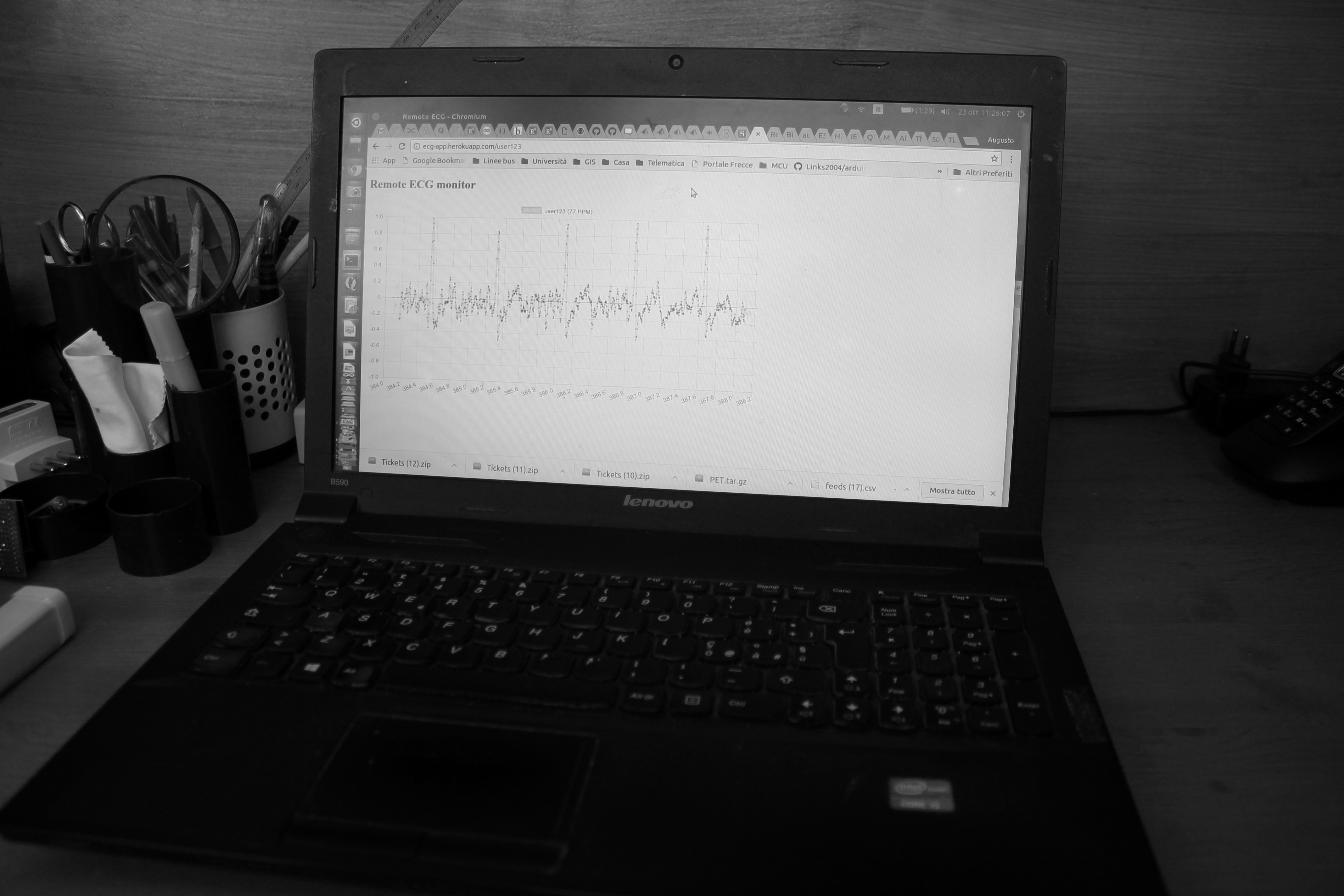}
	\caption{The remote \ac{EKG} display on doctor's side}
	\label{fig:doctor}
\end{figure*}

The interface on doctor's side 
(see figure \ref{fig:doctor}) should be extended with a switch to select filtering, feature extraction and evaluation tools, and the possibility to save, and replay, the trace.  A Smartphone interface is close to reach, but the limited size of the display makes the \ac{EKG} significantly less readable.

\section{Discussion}

An open-source on-line remote \ac{EKG} architecture has been introduced. This kind of service is already offered in return for a payment, as part of health-care packages, and it is usually based on proprietary hardware/software resources. In contrast, this paper aims at an architecture based on open source resources, with an attitude that we summarize in the \ac{OpLoC} acronym: hardware, software, and communication protocols are affordable open source products. A few of them have been designed for the purpose (namely, the client/server software and the serial protocol), but the rule is to use \ac{COTS} resources. One of the resources is in the cloud, the web-server that transfers the \ac{EKG} from the patient to the physician: no exception, it is a {\em plain} open-source web server. To demonstrate that the architecture is feasible and to evaluate its performance, we implemented a prototype that is exhaustively described and reproducible. The \ac{OpLoC} principles foster a wider diffusion of a useful device on a more competitive basis, and make it applicable to disadvantaged or marginal regions.

\section{Abbreviations}

\begin{acronym}
	\acro{OpLoC}{Open Source, Low Cost, COTS}
	\acro{COTS}{Commercial Off-the-shelf}
	\acro{EKG}{electrocardiogram}
	\acro{UA}{User Agent}
	\acro{FQDN}{Fully Qualified Domain Name}
	\acro{HTTP}{Hypertext Transfer Protocol}
	\acro{HTTPS}{Secure Hypertext Transfer Protocol}
	\acro{URL}{Uniform Resource Locator}
	\acro{JSON}{JavaScript Object Notation}
	\acro{WSGI}{Web Server Gateway Interface}
	\acro{MCU}{Micro-Controller Unit}
	\acro{SBC}{Single Board Computer}
	\acro{PaaS}{Platform as a Service}
	\acro{AP}{Access Point}
	\acro{WiFi}{Wireless Fidelity}
	\acro{BAN}{Body (or Personal) Data Network}
	\acro{MSN}{Medical Sensor Network}
	\acro{PC}{Personal Computer}
	\acro{AWS}{Amazon Web Services}
	\acro{NAT}{Network Address Translation}
	\acro{TLS}{Transport Level Security}
	\acro{ADC}{Analog Digital Converter}
	\acro{UART}{Universal Asynchronous Receiver-Transmitter}
	\acro{IP}{Internet Protocol}
	\acro{TCP}{Transport Control Protocol}
\end{acronym}

\section{Bibliography}

\bibliography{biblio,cur,rfc}

\end{document}